\documentclass[12pt]{iopart}
\usepackage{epsf}

\begin{document}

\title[Decoherence of two maximally entangled qubits]
{Decoherence of two maximally entangled qubits in a~lossy
nonlinear cavity}
\author{Adam Miranowicz}
\date{\today}

\address{$^1$ Nonlinear Optics Division, Physics Institute, Adam
 Mickiewicz University, 61-614 Pozna\'n, Poland}
\address{$^2$ CREST Research Team for Interacting Carrier
 Electronics, The Graduate University for Advanced Studies
 (SOKEN-DAI), Hayama, Kanagawa 240-0193, Japan}

\begin{abstract}
Decoherence effect on quantum entanglement of two optical qubits in
a lossy cavity interacting with a nonlinear medium (Kerr
nonlinearity) is analyzed. The qubits are assumed to be initially
in the maximally entangled states (Bell or Bell-like states) or the
maximally entangled mixed states, on the example of Werner and
Werner-like states. Two kinds of measures of the entanglement are
considered: the concurrence to describe a decay of the entanglement
of formation of the qubits, and the negativity to determine a decay
of the entanglement cost under
positive-partial-transpose-preserving operations. It is observed
that the Kerr nonlinearity, in the discussed decoherence model,
does not affect the entanglement of the qubits initially in the
Bell or Werner states, although the evolution of the qubits can
depend on this nonlinearity explicitly. However, it is shown that
for the initial Bell-like state and the corresponding Werner-like
state, the loss of the entanglement can be periodically reduced by
inserting the Kerr nonlinearity in the lossy cavity. Moreover, the
relativity of the entanglement measures is demonstrated, to our
knowledge for the first time, as a result of a physical process.

\vspace{5mm}\noindent PACS numbers: 03.65.Yz, 42.50.Dv, 03.67.Mn


\end{abstract}

\pagenumbering{arabic}

\section{Introduction}

Decoherence, resulting from the unavoidable and irreversible
coupling of a quantum system to its environment, turns a correlated
quantum state of the system into a classical statistical mixture
\cite{Giu96}. Decoherence, causing usually a loss of quantum
entanglement, is one of the major limitations of practical
capabilities of quantum computers \cite{Bra01}. Thus, the analysis
of the dynamics of entangled quantum two-level systems (qubits)
coupled to the environment, represented by a thermal reservoir, is
of a particular importance. In this paper, we will study the loss
of the entanglement due to a dissipative nonlinear interaction of
two optical qubits, which are implemented by superpositions of
vacuum and single-photon states of two cavity modes, and assumed to
be initially in the maximally entangled states (MESs) or the
maximally entangled mixed states (MEMSs).

It is a well accepted fact that there is no unique way to quantify
mixed-state entanglement and thus various measures with different
operational interpretations have been proposed to describe
different aspects of the entanglement. We will apply the
concurrence \cite{Woo98}, a measure related to the entanglement of
formation \cite{Ben96f}, and the negativity
\cite{Per96,Zyc98,Eis99}, a measure corresponding to an
operation-limited entanglement cost \cite{Aud03,Ish04}. We have
chosen these particular measures as they are similar to each other
from a physical point of view and, moreover, can easily be
calculated contrary to other measures including the entanglement of
distillation or the relative entropy of entanglement.

Our analysis is related to a new regime of quantum nonlinear optics
involving highly-efficient nonlinear interactions between very weak
optical fields, which has been recently demonstrated experimentally
in, e.g., dense atomic media by using an electromagnetically
induced transparency (EIT) to resonantly enhance nonlinearities
(for a review see \cite{Luk01}). In particular, observation of
giant Kerr nonlinearities has been predicted \cite{Sch96} and first
measured in an ultracold gas of sodium atoms to be $\sim 10^6$
greater than those in the conventional optical materials
\cite{Hau99}. Physical realizations of a Kerr nonlinear cavity
enabling strong interaction of photons was suggested by
Imamo\v{g}lu et al. \cite{Sch96,Ima97} and then studied by others
\cite{Reb99,Dua00,Hon02,Kua03,Kan03}. Motivated by these advances,
there is an increasing interest to apply the Kerr nonlinearities
for quantum information purposes \cite{kerr-qc}, including the
problem of generation of highly-entangled states (see
\cite{Dua00,kerr-ent} and references therein). Nevertheless, the
effects of decoherence on the entanglement of fields interacting
via the Kerr nonlinearity have not been discussed in greater detail
yet.

The paper is organized as follows. In section 2, we define the
entanglement measures to be used in our description of decoherence.
The model and its solution for two optical qubits in a lossy
nonlinear cavity are presented in section 3. The main results
concerning the decoherence of the qubits being initially in the
maximally entangled states and the maximally entangled mixed states
are presented in sections 4 and 5, respectively. A physical
implementation of the model and discussion of the results are given
in section~6.

\section{Entanglement measures}

We will apply two measures of entanglement to analyze the effect of
decoherence on the entangled qubit states. The first measure is the
concurrence defined for two qubits as~\cite{Woo98}
\begin{equation}
C(\hat{\rho})=\max \{2\max_{i}\lambda _{i}-\sum_{i=1}^{4}\lambda
_{i},0\} \label{N01}
\end{equation}
where $\lambda _{i}$ are the square roots of the eigenvalues of the
matrix $\hat{\rho}(\hat{\sigma}_{1y}\otimes
\hat{\sigma}_{2y})\hat{\rho} ^{\ast }(\hat{\sigma}_{1y}\otimes
\hat{\sigma}_{2y})$ with $\hat{\sigma} _{jy}$ being the Pauli spin
matrix of the $j$th qubit and the asterisk denotes complex
conjugation. The entanglement of formation, $E_{F}(\hat{\rho})$,
which characterizes the amount of entanglement necessary to create
the entangled state \cite{Ben96f}, is for two qubits given by a
simple monotonic function of the concurrence \cite{Woo98}
\begin{equation}
E_{F}(\hat{\rho})= H\{\frac{1}{2}[1+\sqrt{1-C(\hat{\rho})^{2}}]\}
\label{N02}
\end{equation}
where $H\{x\}=-x\log _{2}x-(1-x)\log _{2}(1-x)$ is the binary
entropy. The entanglement of formation and, equivalently, the
concurrence vanish for an unentangled state, and are equal to one
for a maximally entangled state. A measure associated with the
entanglement of formation is the entanglement cost defined as
\cite{Ben96f} $\lim_{n\rightarrow\infty}E_F(\hat{\rho}^{\otimes
n})/n$ which, in general, is quite difficult to calculate. Thus,
for simplicity, we will describe the entanglement cost limited to a
special class of operations to be specified in the following.

Another useful measure of the entanglement is the negativity
\cite{Zyc98,Eis99}, which corresponds to a quantitative version of
the Peres-Horodecki criterion \cite{Per96}. We adopt here the
following definition
\begin{equation}
\mathcal{N}(\hat{\rho})=2\max (0,-\sum_{j}\mu _{j})  \label{N03}
\end{equation}
where the sum is taken over the negative eigenvalues $\mu _{j}$ of
the partial transpose $\hat{\rho}^{T_{A}}$ of the density matrix
$\hat{\rho}$ of the system. For two-qubit pure or mixed states, the
sum in (\ref{N03}) can be skipped as $\hat{\rho}^{T_{A}}$ has at
most one negative eigenvalue \cite{San98}. The negativity satisfies
the standard conditions for a useful measure of the entanglement
\cite{Eis01,Vid02}. For two-qubit states, the negativity, defined
by (\ref{N03}), becomes 1 for a MES and vanishes for an unentangled
state, the same as the concurrence. Recently, Audenaert et al.
\cite{Aud03} and supplementary Ishizaka \cite{Ish04} have provided
an operational interpretation of the logarithmic negativity,
defined by \cite{Vid02}
\begin{equation}
E_N(\hat{\rho})=\log _{2}[{N} ({\hat{\rho} })+1], \label{N04}
\end{equation}
as a measure of the entanglement cost for the exact preparation of
a two-qubit quantum state $\hat{\rho}$ under quantum operations
preserving the positivity of the partial transpose (PPT).

For an arbitrary two-qubit pure state
\begin{equation}
|\Psi \rangle =c_{00}|00\rangle +c_{01}|01\rangle
+c_{10}|10\rangle +c_{11}|11\rangle \label{N05}
\end{equation}
with the normalized complex amplitudes $c_{ij}$, we have the
concurrence and negativity equal to each other and given by a
simple formula
\begin{equation}
\mathcal{N}_{\Psi }=C_{\Psi }=2|c_{00}c_{11}-c_{01}c_{10}|.
\label{N06}
\end{equation}
However, for qubits in a mixed state, the entanglement measures are
usually different. In general, the inequality ${\cal
N}(\hat{\rho})\le C(\hat{\rho})$ holds for an arbitrary two-qubit
state $\hat{\rho}$ as first observed by numerical investigation by
Eisert and Plenio \cite{Eis99} and \.Zyczkowski \cite{Zyc99}, and
then proved by Verstraete et al. \cite{Ver01a},

Eisert and Plenio \cite{Eis99} raised an intriguing problem of the
relativity of entanglement measures: if according to one measure of
the entanglement the state $\hat{\rho}_1$ is more entangled than
$\hat{\rho}_2$ then does it imply that $\hat{\rho}_1$ is also more
entangled than $\hat{\rho}_2$ according to another entanglement
measure? By Monte Carlo simulation of thousands of two-qubit
states, it was observed that indeed the condition
\begin{equation}
C(\hat{\rho}_1) < C(\hat{\rho}_2) \Leftrightarrow {\cal
N}(\hat{\rho}_1) < {\cal N}(\hat{\rho}_2) \label{N07}
\end{equation}
can be violated by some states $\hat{\rho}_1$ and $\hat{\rho}_2$
\cite{Eis99,Zyc99,Wei03}. It should be stressed that this odd
looking property is physically sound since such incomparable states
$\hat{\rho}_1$ and $\hat{\rho}_2$ cannot be transformed into each
other with unit efficiency by any local quantum operations and
classical communication (LQCC). In general, all good asymptotic
entanglement measures are either identical or put different
orderings of quantum states as implied by the requirements of
equivalence and continuity of the measures on pure states
\cite{Vir00}. Thus, by comparing various entanglement measures
defined to examine different methods of the entanglement
preparation and/or its use, one indeed can arrive at the problem of
different state orderings imposed by the measures. The only
alternative to avoid the state-ordering ambiguity is to declare one
entanglement measure for mixed states as the unique one, but this
would preclude us from examining the problems of how to prepare the
entanglement and how to make use of it \cite{Par02}. Here, we will
give simple analytical examples of states differently ordered by
the concurrence and negativity, thus we will explicitly demonstrate
the relativity of these entanglement measures.

\section{Model and its solution}

Decoherence effects on optical modes (qubits) in a lossy nonlinear
cavity can be described by a model of $N$ coupled dissipative
nonlinear oscillators represented by the following prototype
Hamiltonian \cite{Per94}\footnote{Hamiltonian $\hat{H}_{\rm NL}$ is
sometimes defined in the normal-ordered form of $\hat{a}_{j}^{\dag
}$ and $\hat{a}_j$. However, such Hamiltonian differs from ours
only in terms proportional to $\chi _{jj}\hat{a}_{j}^{\dag
}\hat{a}_j$, which can be incorporated in the free Hamiltonian
$\hat{H}_0$, so this modification does not effect the
entanglement.}
\begin{equation}
\hat{H}=\hat{H}_{0}+\hat{H}_{\rm NL}+\hat{H}_{I} \label{N08}
\end{equation}
where
\begin{eqnarray}
\hat{H}_{0} &=&\hbar \sum_{j=1}^{N}\omega _{j}\hat{a}_{j}^{\dag
}\hat{a} _{j}+\hbar \sum_{k}\sum_{j=1}^{N}\Omega
_{k}^{(j)}(\hat{b}_{k}^{(j)})^{\dag }
\hat{b}_{k}^{(j)},  \label{N09} \\
\hat{H}_{\rm NL} &=&\hbar \sum_{i,j=1}^{N}\chi
_{ij}\hat{a}_{i}^{\dag }\hat{a}
_{i}\hat{a}_{j}^{\dag }\hat{a}_{j}, \label{N10}\\
\hat{H}_{I} &=&\hbar
\sum_{k}\sum_{j=1}^{N}[g_{k}^{(j)}\hat{a}_{j}^{\dag }
\hat{b}_{k}^{(j)}+{\rm h.c.}] \label{N11}
\end{eqnarray}
and $\hat{a}_{j}$ is the annihilation operator for the $j$th system
oscillator at the frequency $\omega _{j}$; $\hat{b}_{k}^{(j)}$ is
the annihilation operator for the $k$th oscillator in the $j$th
reservoir at the frequency $\Omega _{k}^{(j)}$; $\chi _{ij}$ are
the nonlinear self-coupling (for $i=j$) and cross-coupling (for
$i\neq j)$ constants proportional to a third-order susceptibility
of the Kerr nonlinear medium (see Sect. 6), and $g_{k}^{(j)}$ are
the coupling constants of the reservoir oscillators. Dissipation of
the system is modelled by its coupling to reservoirs of oscillators
as described by Hamiltonian $\hat{H}_{I}$. The evolution of the
dissipative system under the Markov approximation is governed by
the following master equation for the reduced density operator
$\hat{\rho}$ in the interaction picture
\begin{eqnarray} \frac{\partial }{\partial t}\hat{\rho}
&=&\frac{1}{i\hbar }[\hat{H}_{\rm NL},
\hat{\rho}]+\sum_{j=1}^{N}\frac{\gamma
_{j}}{2}\{\bar{n}_{j}(2\hat{a}_j^{\dag }
\hat{\rho}\hat{a}_j-\hat{a}_j\hat{a}_j^{\dag
}\hat{\rho}-\hat{\rho}\hat{a}_j\hat{a}_j
^{\dag })  \nonumber \\
&&\hspace{3cm}+(\bar{n}_{j}+1)(2\hat{a}_j\hat{\rho}\hat{a}_j^{\dag
}-\hat{a}_j^{\dag }\hat{a}_j \hat{\rho}-\hat{\rho}\hat{a}_j^{\dag
}\hat{a}_j)\} \label{N12}
\end{eqnarray}
where $\bar{n}_{j}$ are the mean thermal occupation numbers and
$\gamma _{j}$ are the damping constants, which will be assumed the
same, $\gamma_1=\gamma_2\equiv \gamma$, in the next sections. With
the help of a disentangling theorem for SU(1,1) in thermofield
dynamics notation, Chaturvedi and Srinivasan \cite{Cha91} have
found a general solution of the master equation (\ref{N12}) both
for the quiet ($\bar{n}_j=0$) and noisy ($\bar{n}_j>0$)
reservoirs. By confining our analysis to the case of only two
oscillators ($N=2$) coupled to the quiet reservoirs, the
Chaturvedi-Srinivasan solution for the density matrix elements
$\langle m_{1},m_{2} |\hat{\rho}(t) |n_{1},n_{2}\rangle$ in the
photon-number basis can be written as
\begin{eqnarray}
\hspace{-1.5cm} \langle
m_{1},m_{2}|\hat{\rho}(t)|n_{1},n_{2}\rangle
=\sum_{p_{1}=0}^{\infty }\sum_{p_{2}=0}^{\infty }R_{1}R_{2}
\langle m_{1}+p_{1},m_{2}+p_{2}|\hat{\rho}
(0)|n_{1}+p_{1},n_{2}+p_{2}\rangle \label{N13}
\end{eqnarray}
where
\begin{eqnarray}
R_{j} &\equiv& R_{j}(m_{j},n_{j},p_{j})=
\left[{{m_j+p_j}\choose{p_j}}{{n_j+p_j}\choose{p_j}} \right]^{1/2}
\left( \frac{\gamma _{j}}{x_{j}}[1-\exp (-x_{j}t)]\right) ^{p_{j}}
\nonumber \\
&&\times \exp \left\{i(\chi _{j1}+\chi _{j2})(m_{j}-n_{j})t
-[x_{j}\ (m_{j}+n_{j}+1)-\gamma _{j}]\frac{t}{2} \right\}
\label{N14}
\end{eqnarray}
with $x_{j}\ =\gamma _{j}+2i[\chi _{j1}(m_{1}-n_{1})+\chi
_{j2}(m_{2}-n_{2})],$ and  $q \choose p$ are binomial coefficients.
In our scheme, qubits can be represented by the single-cavity modes
restricted in the Hilbert space spanned by vacuum and single-photon
states (see, e.g., \cite{Gio00}). Then, for the qubit states,
solution (\ref{N13}) simplifies to the summations over
$p_1,p_2=0,1$ only.

By assuming no dissipation ($\gamma_1=\gamma_2=0$) in our system,
the evolution is governed by the unitary operator
$\exp(-i\hat{H}_{\rm NL}t/\hbar )$. It implies that, for the two
qubits initially in a pure state (\ref{N05}), the concurrence and
negativity evolve periodically as follows:
\begin{eqnarray}
\hspace{-1cm} \mathcal{N}_{\Psi}(\gamma=0,t)=C_{\Psi}(\gamma=0,t)
=2|\exp (2i\chi _{12}t)c_{00}(0)c_{11}(0)-c_{01}(0)c_{10}(0)|
\label{N15}
\end{eqnarray}
depending on the cross-coupling $\chi _{12}$ but not on the
self-coupling constants $\chi _{11}$ and $\chi _{22}$. One can
observe that the evolution of the qubits in the nonlinear medium
can lead to a periodical generation of entangled states. Even for
the initial separable state
\begin{equation}
|\Psi(0)\rangle =(d_{1}|0\rangle _{1}+d_{2}|1\rangle _{1})\otimes
(d_{3}|0\rangle _{2}+d_{4}|1\rangle _{2}), \label{N16}
\end{equation}
where $|d_1|^2+|d_2|^2=|d_3|^2+|d_4|^2=1$ and none of the
amplitudes $d_{i}$ is zero, the concurrence and negativity
periodically become positive
\begin{eqnarray}
\mathcal{N}_{\Psi}(\gamma=0,t)=C_{\Psi}(\gamma=0,t) =
4|d_{1}d_{2}d_{3}d_{4}\sin (\chi _{12}t)| \label{N17}
\end{eqnarray}
which corresponds to the entanglement of up to
$H\{\frac{1}{2}[1+\sqrt{1-16|d_{1}d_{2}d_{3}d_{4}|^{2}}]\}$ ebits.
In particular, the initial state (\ref{N16}) with all the
amplitudes equal to $1/\sqrt{2}$, i.e.,
\begin{equation}
|\Psi(0)\rangle =\frac{|0\rangle _{1}+|1\rangle
_{1}}{\sqrt{2}}\otimes \frac{|0\rangle _{2}+|1\rangle
_{2}}{\sqrt{2}} \equiv |+,+\rangle, \label{N18}
\end{equation}
evolves into a maximally entangled state, defined below by
(\ref{N26}), having exactly 1 ebit at the evolution moments
$t=(1+2n)\pi/(2\chi_{12})$ ($n=0,1,...$). Nevertheless, the MESs
are not generated if our system is subjected to dissipation.

\section{Decoherence of the maximally entangled states}

Let us assume that two qubits are initially in the Bell states
\begin{equation}
|\psi _{\pm }\rangle =\frac{1}{\sqrt{2}}(|01\rangle \pm |10\rangle
) \label{N19}
\end{equation}
which evolve in the dissipative nonlinear cavity into the
following mixed states
\begin{eqnarray}
\hat{\rho} _{\psi \pm }(t) &=&\frac{1}{2}\{2(1-g)|00\rangle
\langle 00|+g(|01\rangle \langle 01|+|10\rangle \langle 10|) \nonumber \\
&&\pm g (e^{i(\chi _{1}-\chi _{2})t}|01\rangle \langle 10|+{\rm
h.c.})\} \label{N20}
\end{eqnarray}
where $g=e^{-\gamma t}$ and $\chi _{i}\equiv\chi _{ii}$. The
evolution is independent of the nonlinear cross-coupling $\chi
_{12}$ but depends on the self-couplings $\chi _{1}$ and $\chi
_{2}$. We find that the concurrence is simply given by
\begin{equation}
C_{\psi }(t)=g \label{N21}
\end{equation}
and the negativity is
\begin{equation}
\mathcal{N}_{\psi }(t)=\sqrt{2g^2-2g+1}+g-1 \label{N22}
\end{equation}
being independent of the sign in (\ref{N19}). As implied by the
form of the density matrices (\ref{N20}), the entanglement
measures are independent of any Kerr couplings. On the other hand,
the initial Bell states
\begin{equation}
|\phi _{\pm }\rangle =\frac{1}{\sqrt{2}}(|00\rangle \pm |11\rangle
) \label{N23}
\end{equation}
evolve in our lossy system into
\begin{eqnarray}
\hat{\rho} _{\phi \pm }(t) &=&\frac{1}{2}\{(2-2g+g^2)|00\rangle
\langle 00|+(1-g)g(|01\rangle \langle 01|
+|10\rangle \langle 10|) \nonumber \\
&&\pm g(e^{i(\chi _{1}+2\chi _{12}+\chi _{2})t}|00\rangle \langle
11|+{\rm h.c.}) +g^2|11\rangle \langle 11|\}. \label{N24}
\end{eqnarray}
Contrary to $\hat{\rho} _{\psi \pm }(t)$, the density matrices
$\hat{\rho} _{\phi \pm }(t) $ depend on the cross-coupling between
the qubits. We find that the concurrence and negativity are the
same for any evolution times and any sign in (\ref{N23}) as given
by
\begin{eqnarray}
C_{\phi }(t)= \mathcal{N}_{\phi }(t)&=&g^2 \label{N25}
\end{eqnarray}
in contrast to $C_{\psi }(t)$ and $\mathcal{N}_{\psi }(t)$, given
by (\ref{N21}) and (\ref{N22}), respectively, which are the same at
$t=0$ and $t=\infty $ only. There are the following important
properties of the discussed entanglement decays. First, the
concurrence (negativity) for the initial Bell states $|\psi _{\pm
}\rangle $ decays slower (faster) than that for $|\phi _{\pm
}\rangle $, as it holds $C_{\psi}(t)>C_{\phi}(t)$ and
$\mathcal{N}_{\psi }(t)< \mathcal{N}_{\phi }(t)$ for any damping
constants $\gamma_k>0$ ($k=1,2$) and any moments of time
$0<t<\infty$. Thus, we provide an explicit example of states
violating condition (\ref{N07}). Second, contrary to the density
matrices, the entanglement measures are independent of the
nonlinear couplings for the initial Bell states (\ref{N19}) and
(\ref{N23}). Thus, decoherence-free evolution in the nonlinear
cavity does not change the entanglement, i.e., $ C_{\psi }(\gamma
=0,t)=\mathcal{N}_{\psi }(\gamma =0,t)=1$. Now we will give an
example of a maximally entangled two-qubit state evolving in the
Kerr medium in such a way that the entanglement depends on the
cross-coupling $\chi _{12}$. Let us analyze the following initial
state
\begin{equation}
|\varphi\rangle =\frac{1}{2}(|00\rangle +|01\rangle +|10\rangle
-|11\rangle )\equiv \frac{1}{\sqrt{2}}(|0,+\rangle +|1,-\rangle )
\label{N26}
\end{equation}
where $|\pm \rangle =(|0\rangle \pm |1\rangle )/\sqrt{2}$. The
state $ |\varphi \rangle $ is a MES since its concurrence and
negativity are equal to one. For brevity, we neglect the
self-couplings, $\chi _{1}=\chi _{2}=0$, which do not affect the
entanglement. Then the initial pure state $|\varphi \rangle$
evolves in the Kerr medium into the mixed state described by
\begin{equation} \hat{\rho}_{\varphi }(t)=\frac{1}{4}\left[
\begin{array}{cccc}
(2-g)^{2} & h\sqrt{g} & h\sqrt{g} & -fg \\
h^{\ast }\sqrt{g} & g(2-g) & g & -fg^{3/2} \\
h^{\ast }\sqrt{g} & g & g(2-g) & -fg^{3/2} \\
-f^{\ast }g & -f^{\ast }g^{3/2} & -f^{\ast }g^{3/2} & g^{2}
\end{array}
\right]   \label{N27}
\end{equation}
given, as usual, in the computational basis $\{|00\rangle
,|01\rangle, |10\rangle ,|11\rangle \}$, and $h=(\gamma fg-2i\chi
_{12})/(\gamma -2i\chi _{12})$, $g=\exp (-\gamma t)$, and $f=\exp
(2i\chi _{12}t)$. If we moreover assume no losses in the nonlinear
cavity ($\gamma =0$), then the evolution of the initial state
$|\varphi \rangle $ results in the entanglement oscillations
described simply by
\begin{equation}
C_{\varphi }(\gamma =0,t)=\mathcal{N}_{\varphi }(\gamma =0,t)=|\cos (\chi
_{12}t)|,  \label{N28}
\end{equation}
which is in contrast to the case of the initial Bell states $|\phi
_{\pm }\rangle $ and $|\psi _{\pm }\rangle $, which evolve without
changing their entanglements. The aperiodic decay of the
entanglement for the density matrix (\ref{N27}) occurs only if
there is no interaction between the qubits and then it is
described by
\begin{equation}
C_{\varphi }(\chi _{12}=0,t)=\frac{1}{2}g(1+g),  \label{N29}
\end{equation}
\begin{equation}
\mathcal{N}_{\varphi }(\chi _{12}=0,t)=\sqrt{x^{2}-4x+1}+g-1
\label{N30}
\end{equation}
where $x=g(1-g)/2$. For nonzero damping and cross-coupling
parameters, both the concurrence $C_{\varphi}(t)$ and the
negativity $\mathcal{N}_{\varphi }(t)$ exhibit decaying
oscillations, as shown by curves (c) in figures 1 and 2,
respectively. The expressions for $C_{\varphi }(t)$ and
$\mathcal{N} _{\varphi }(t)$ are quite lengthy in general case of
non-zero $\chi _{12}$ and $\gamma$, so we do not present them
here. Instead we give approximate formulas for the envelope
function of the concurrence
\begin{eqnarray}
{C}_{\varphi }^{{\rm env}}(t) \approx
\frac{g}{4}[\sqrt{x-\textstyle{
\frac{2}{3}}(z+\sqrt{2(x-2\,y)(x+y-z)})} +g-1] \label{N31}
\end{eqnarray}
where $x=27-14g+3g^{2}$, $y=\sqrt{159-129g+37g^{2}-3g^{3}}$, and
$z=\sqrt{ (x+y)^{2}-9y^{2}}$, and for the envelope of the
negativity
\begin{equation}
\mathcal{N}_{\varphi }^{{\rm env}}(t)\approx \frac{1}{6}\left(
2\mathrm{Re} \sqrt[3]{v+i3(1-g)g\sqrt{3w}}-(2-g)^{2}-g\right)
\label{N32}
\end{equation}
where $v=8g^{6}-18g^{5}-93g^{4}+324g^{3}-273g^{2}+180g-64$ and $
w=116g^{6}-316g^{5}+297g^{4}+930g^{3}-515g^{2}+624g+16$. Equation
(\ref{N32}) was derived by assuming only that the cross-coupling
$\chi _{12}$ is much stronger than the damping constant $\gamma $,
which implies that the function $h$ in the density matrix
(\ref{N27}) approaches unity. The envelope functions, given by
(\ref{N31}) and (\ref{N32}), are depicted by curves (e) in figures
1 and 2, respectively. Another simpler but far less accurate
approximation of the negativity envelope function can be given by
\begin{equation}
\mathcal{N}_{\varphi }^{{\rm env}}(t)\approx
\frac{1}{2}\sqrt{\frac{ g^{3}(g^{3}-3g^{2}-g+11)}{g^{2}-3g+4}}
\label{N33}
\end{equation}
which was obtained by using the general properties of the
eigenvalues $\mu_i$ of the partial transpose $\hat{\rho}^{T_1}
_{\varphi }(t)$ of density matrix (\ref{N27}), including $\sum_i
\mu_i=1$ and $\prod_i \mu_i=\det \hat{\rho}^{T_1} _{\varphi }(t)$,
and observing that there exist two eigenvalues $\mu_i$ summing up
approximately to zero. It is worth noting that the envelope
functions, the same as those given by (\ref{N31}) and (\ref{N32}),
are for the system initially in the separable state given by
(\ref{N18}), for which $\hat{\rho}(t)$ has the form of (\ref{N27})
but with the functions $h$ and $f$ modified as follows
$h=[\gamma(2+fg)-2i\chi _{12}]/(\gamma -2i\chi _{12})$ and
$f=-\exp (2i\chi _{12}t)$. Note also that the envelope functions
(\ref{N31})--(\ref{N33}) are independent of the cross-coupling
$\chi _{12}$ under assumption $ \chi _{12}\gg \gamma $ but, even
in this regime, the period of entanglement oscillations is a
function of $\chi _{12}$. A closer comparison of the entanglement
for the Kerr interacting and non-interacting qubits in the lossy
cavity leads us to the following inequalities $ {C}_{\varphi
}^{{\rm env}}(t_{n}) = C_{\varphi }(\chi _{12}>0,t_{n})>{C}
_{\varphi }(\chi _{12}=0,t_n)$ and $\mathcal{N}_{\varphi }^{{\rm
env}}(t_{n}) =\mathcal{N}_{\varphi }(\chi
_{12}>0,t_{n})>\mathcal{N}_{\varphi }(\chi _{12}=0,t_n) $ valid
for the moments of time equal to $t_{n}=n\pi /\chi _{12}$ for
$n=1,\cdots $. By comparing the entanglement measures for the all
analyzed MESs (see figures 1 and 2) we can finally conclude that
\begin{eqnarray}
C_{\psi}(t)\ge  C_{\varphi }^{\rm env}(\chi _{12}\ge 0,t) \ge
C_{\varphi }(\chi
_{12}=0,t) \ge  C_{\phi}(t),\label{N34} \\
\mathcal{N}_{\psi}(t)\le  \mathcal{N}_{\varphi }(\chi
_{12}=0,t)\le \mathcal{N}_{\phi}(t) \le  \mathcal{N}^{\rm
env}_{\varphi}(\chi _{12} \ll\!\!\!\!\!\! / \,  \gamma,t)
\label{N35}
\end{eqnarray}
where the equalities hold for the nonzero damping constant $\gamma$
at the evolution moments $t=0$ and $t=\infty$, while for $\gamma=0$
at any times $t$. Inequalities (\ref{N34})--(\ref{N35}), except
those for $C_{\varphi }^{\rm env}(\chi _{12}\ge 0,t)$ and
$\mathcal{N}^{\rm env}_{\varphi}(\chi _{12} \ll\!\!\!\!\!\! / \,
\gamma,t)$, can be proved analytically, while the remaining
inequalities were checked numerically for a large class of
parameters. Note that for small values of $\chi _{12}$ in
comparison to $\gamma$ it holds $\mathcal{N}_{\varphi }(\chi
_{12},t)\le \mathcal{N} _{\phi}(t)$, nevertheless the last
inequality in (\ref{N35}) is satisfied even if $\chi _{12} \sim
\gamma$, and more pronounced for $\chi _{12} \gg \gamma$ (see
figure 2), which is the condition assumed in the derivation of
(\ref{N31})--(\ref{N33}). Obviously, inequalities corresponding to
(\ref{N34}) hold for the entanglement of formation, $E_{F}(t)$, and
those corresponding to (\ref{N35}) are also valid for the
PPT-entanglement cost, $E_{N}(t)$. The main conclusion is the
following physical interpretation of inequalities
(\ref{N34})--(\ref{N35}): by enabling Kerr interactions between the
qubits initially in the Bell-like state, given by (\ref{N26}), the
loss of the entanglement can be periodically reduced.

\begin{figure}
\epsfxsize=12cm\epsfbox{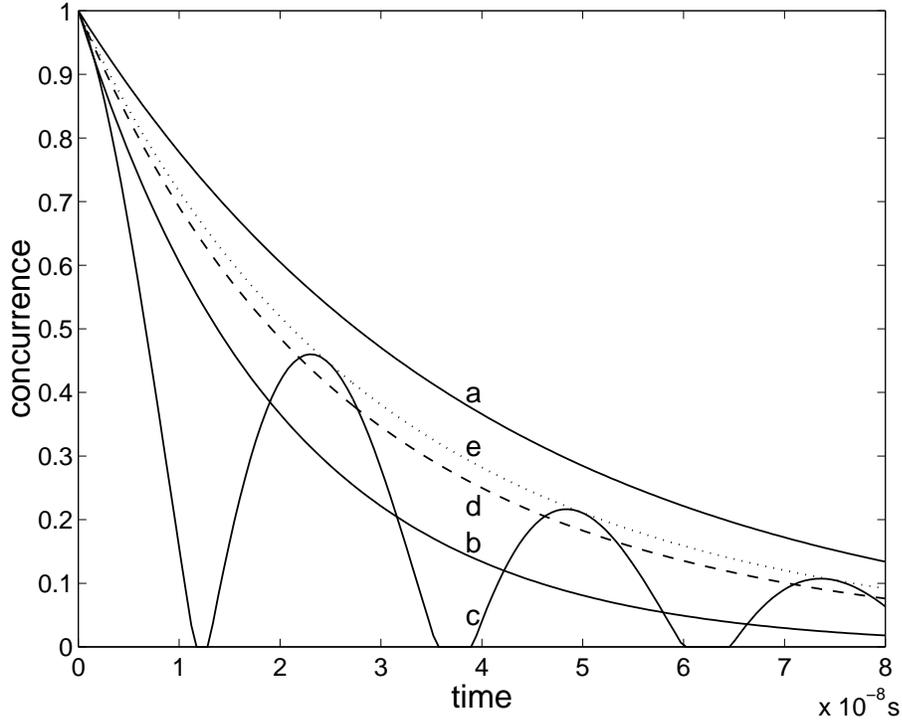} \caption{Decay of the concurrence
for the initial Bell and Bell-like states: (a) $C_{\psi}(t)$, (b)
$C_{\phi}(t)$, (c) $C_{\varphi}(\chi'_{12},t)$, (d)
$C_{\varphi}(\chi_{12}=0,t)$ (dashed curve), and (e) $C^{\rm env }
_{\varphi}(\chi_{12}',t)$ (dotted curve) for cross-coupling
constant $\chi'_{12}=20$~rad~MHz and damping constant
$\gamma=4$~rad~MHz.}
\end{figure}

\begin{figure}
\epsfxsize=12cm\epsfbox{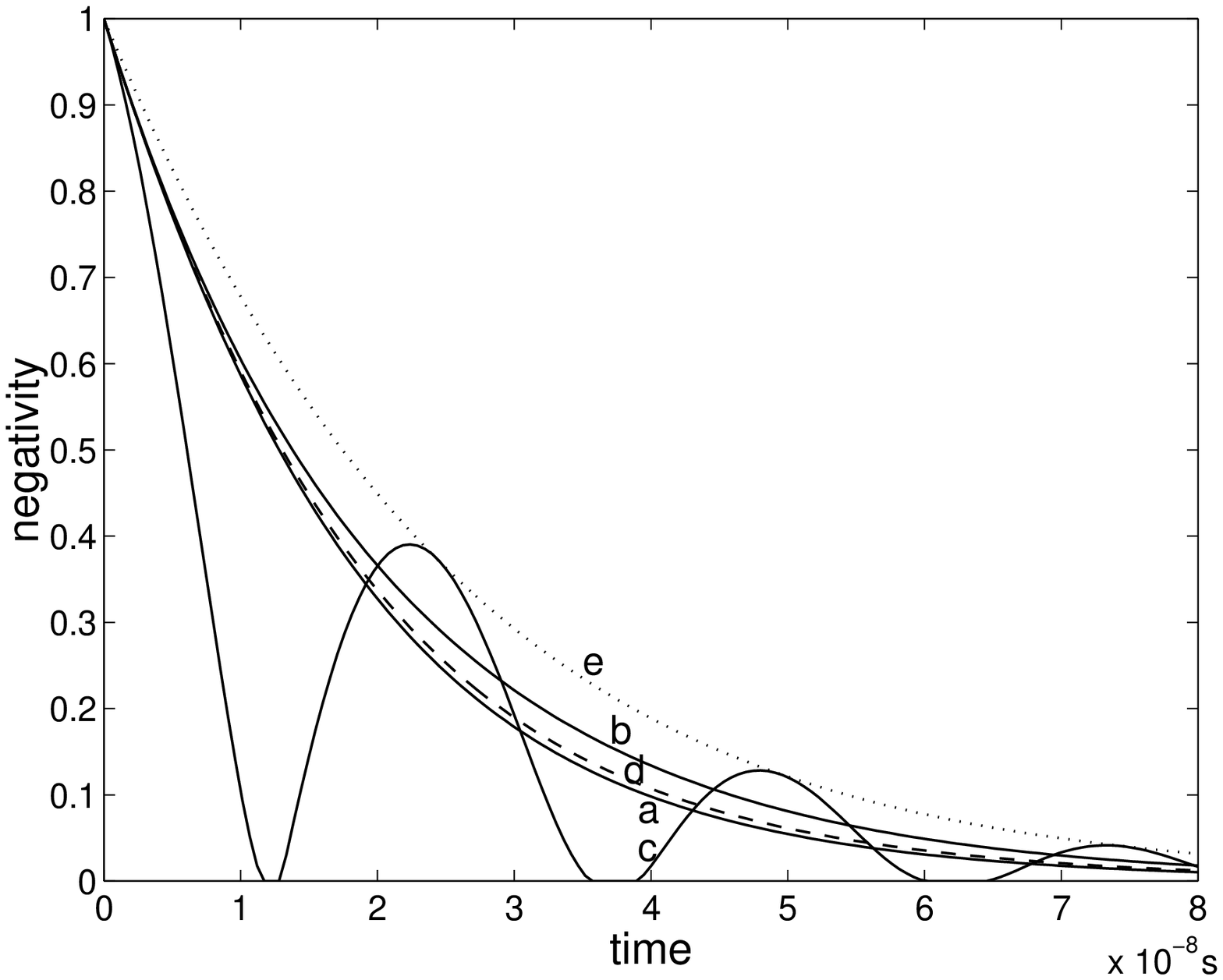} \caption{Decay of the negativity
for the same Bell(-like) states and interactions as in figure 1:
(a) $\mathcal{N} _{\psi}(t)$, (b) $\mathcal{N}_{\phi}(t)$, (c)
$\mathcal{N}_{\varphi} (\chi'_{12},t)$, (d) $\mathcal{N}_{\varphi}
(\chi_{12}=0,t)$, and (e) $\mathcal{N}^{\rm env }
_{\varphi}(\chi'_{12},t)$. }
\end{figure}

\begin{figure}
\hspace*{0mm} \epsfxsize=7cm\epsfbox{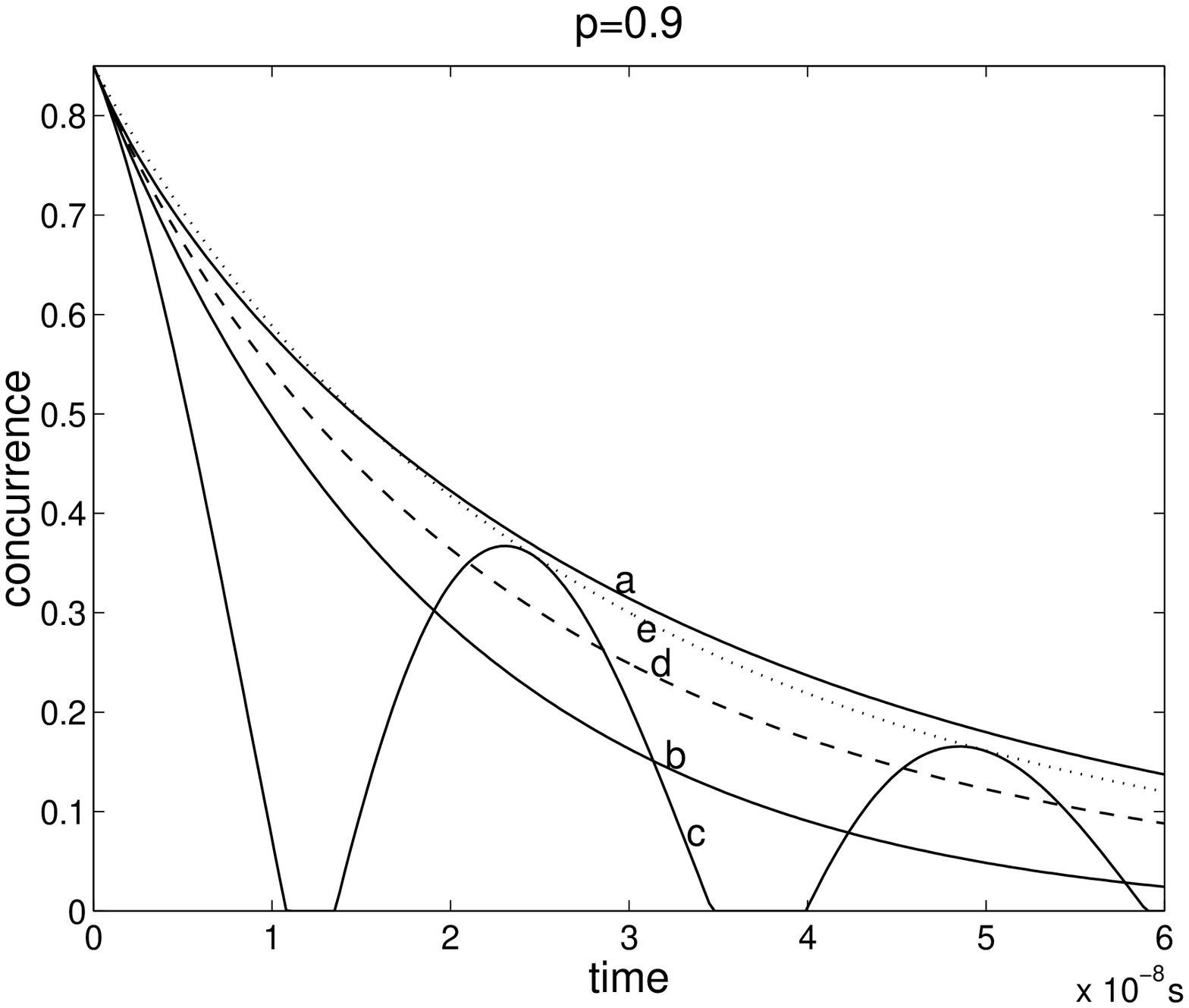} \hspace*{-0.0cm}
\epsfxsize=7cm\epsfbox{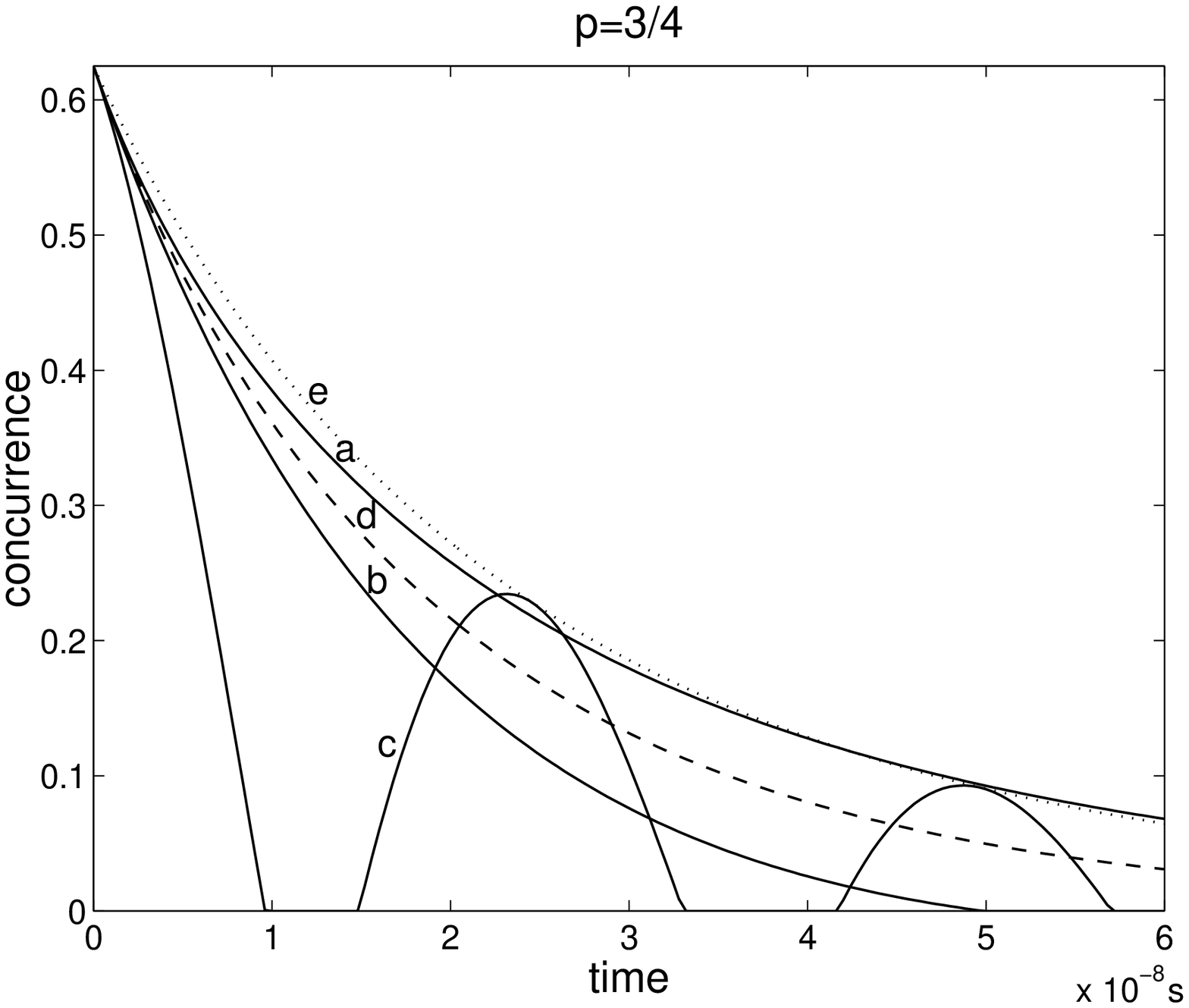} \vspace{0mm}

\hspace*{0mm} \epsfxsize=7cm\epsfbox{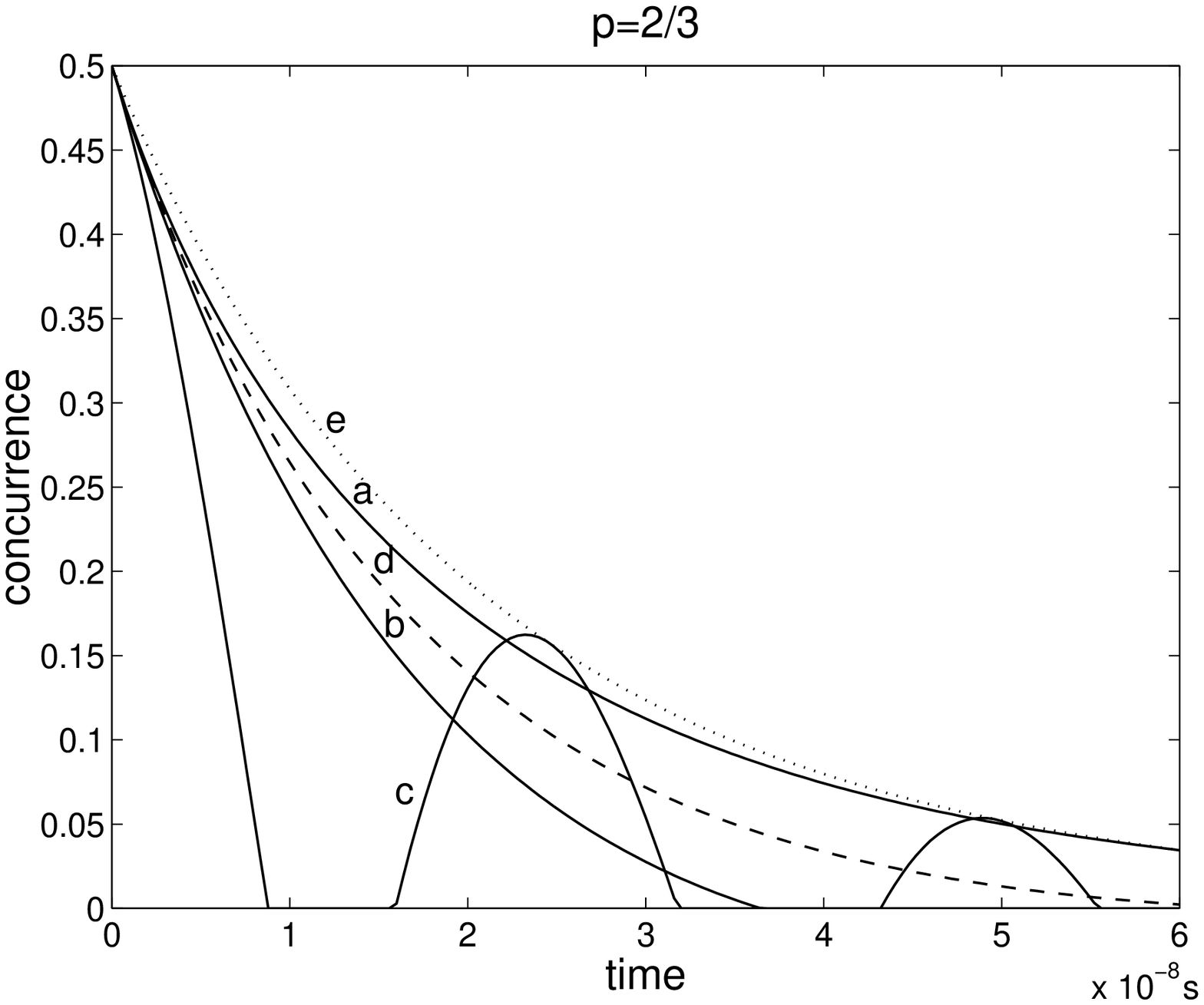} \hspace*{-0.0cm}
\epsfxsize=7cm\epsfbox{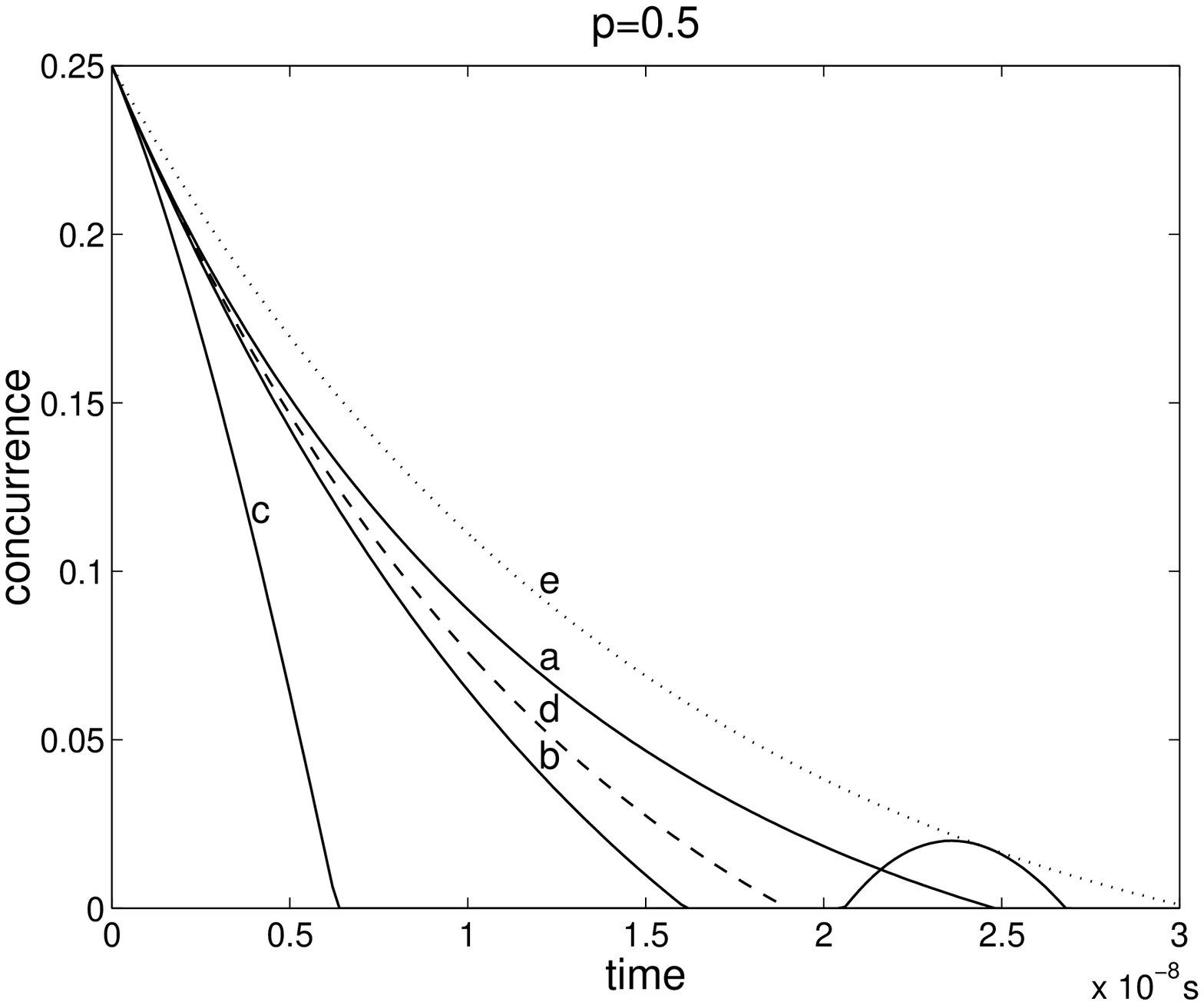} \vspace{0mm}  \caption{Decay of
the concurrence for the initial Werner(-like) states: (a)
$C_{p\psi}(t)$, (b) $C_{p\phi}(t)$, (c)
$C_{p\varphi}(\chi'_{12},t)$, (d) $C_{p\varphi}(\chi_{12}=0,t)$,
and (e) $C^{\rm env } _{p\varphi}(\chi'_{12},t)$ for various values
of parameter $p$; $\gamma$ and $\chi'_{12}$ are the same as in
figure 1.}
\end{figure}

\begin{figure}
\hspace*{0mm} \epsfxsize=7cm\epsfbox{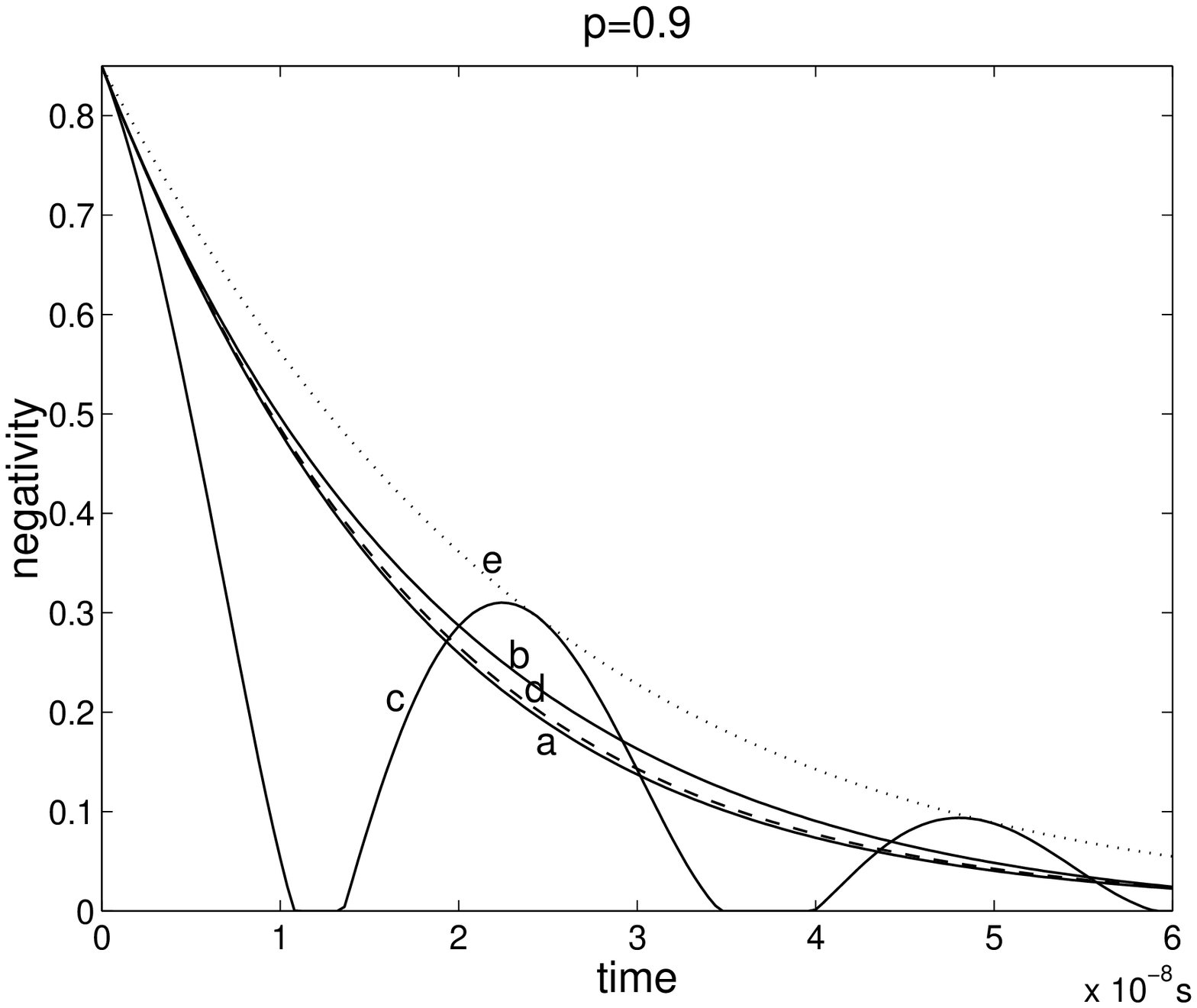} \hspace*{-0.0cm}
\epsfxsize=7cm\epsfbox{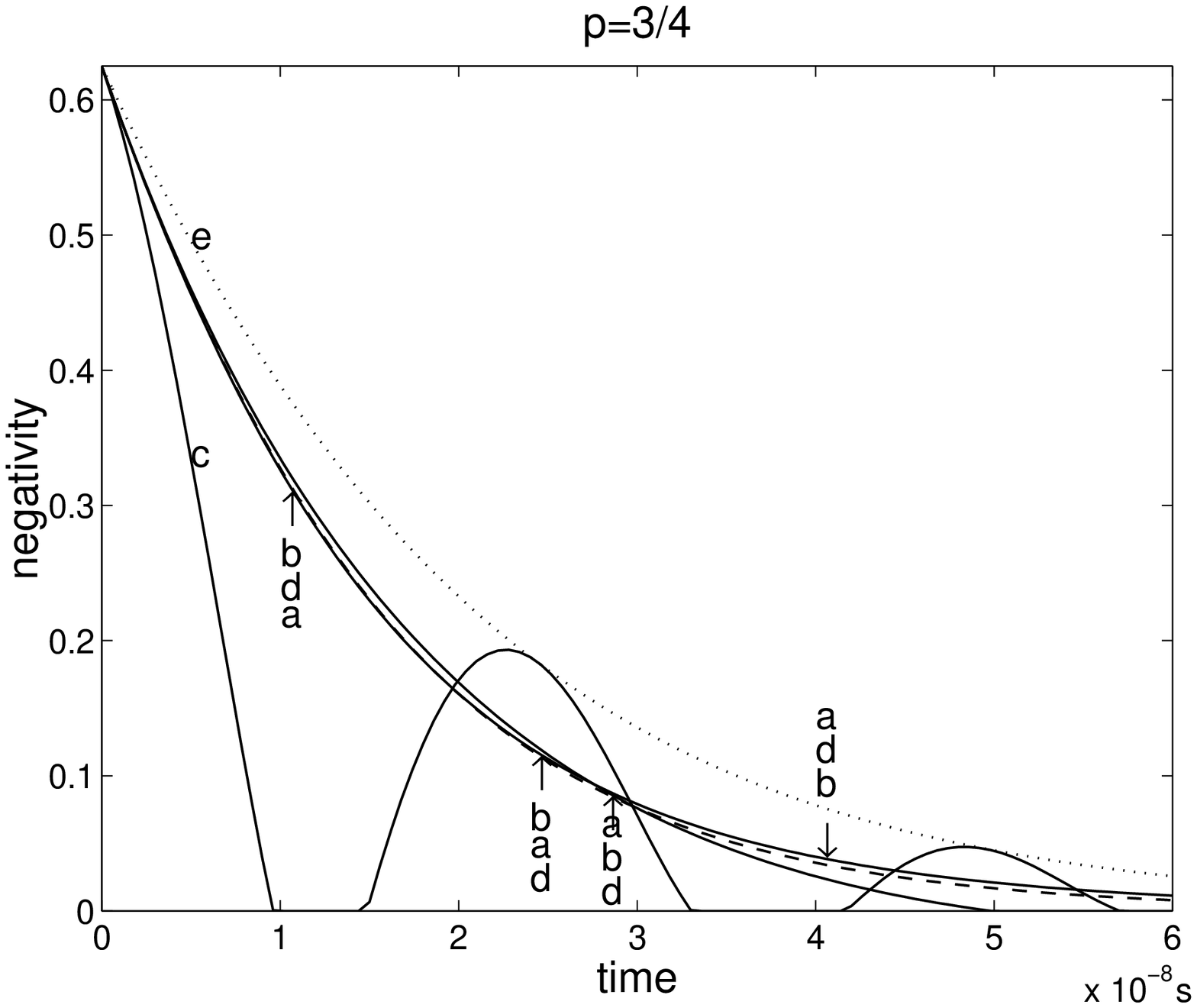} \vspace{0mm}

\hspace*{0mm} \epsfxsize=7cm\epsfbox{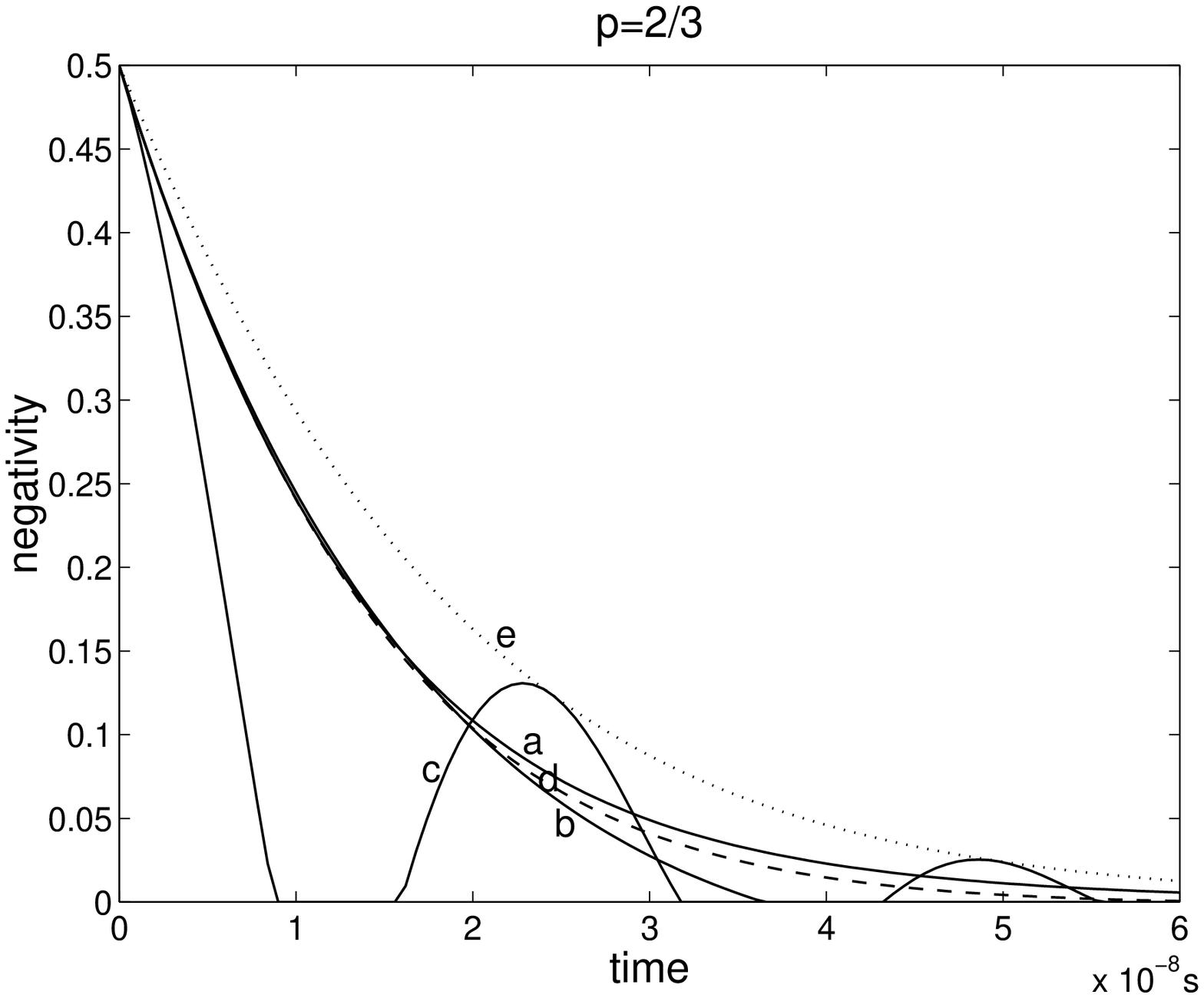} \hspace*{-0.0cm}
\epsfxsize=7cm\epsfbox{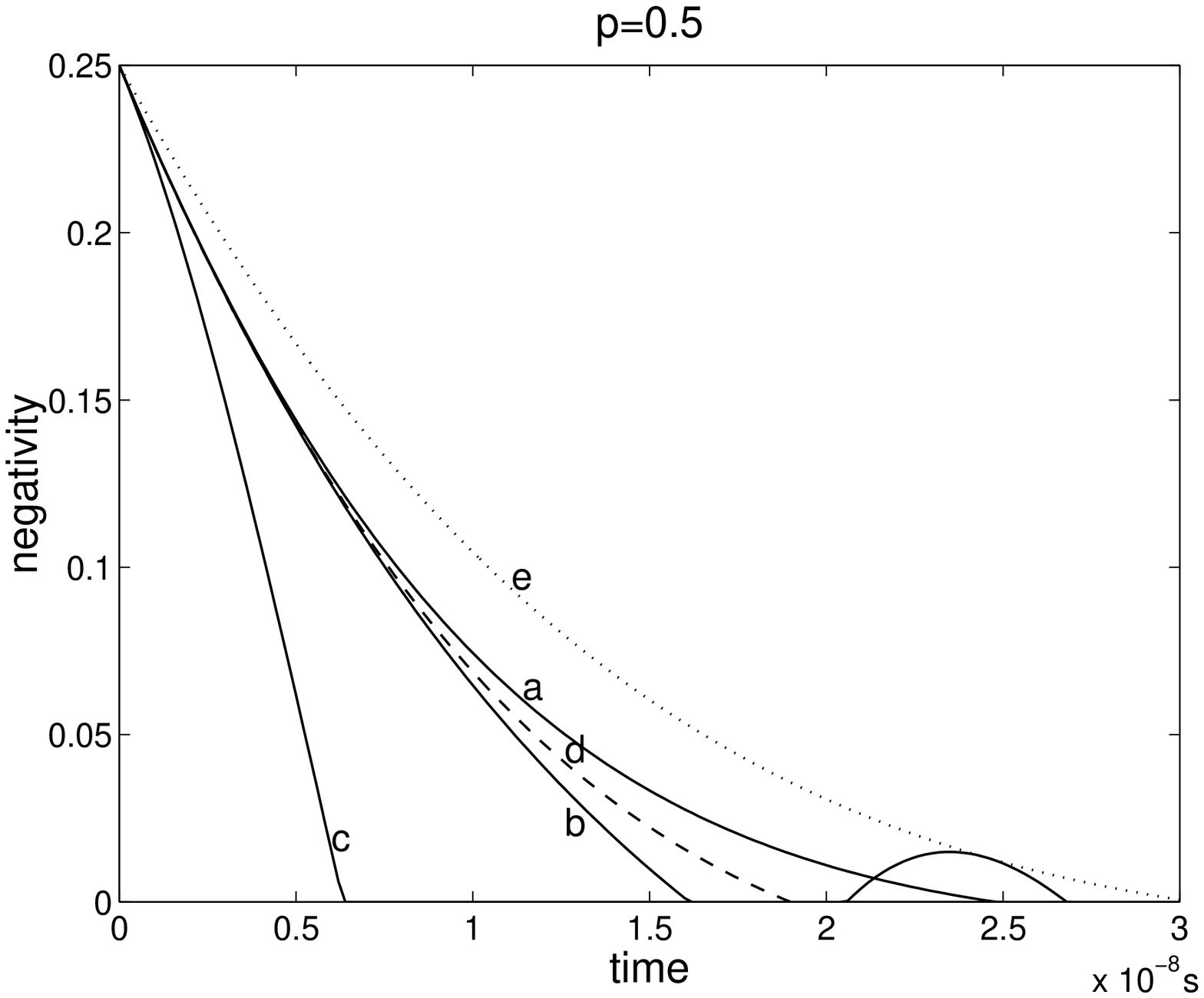} \vspace{0mm}  \caption{Decay of
the negativity for the same Werner(-like) states and interactions
as in figure 3: (a) $\mathcal{N} _{p\psi}(t)$, (b)
$\mathcal{N}_{p\phi}(t)$, (c) $\mathcal{N}_{p\varphi}
(\chi'_{12},t)$, (d) $\mathcal{N}_{p\varphi} (\chi_{12}=0,t)$, and
(e) $\mathcal{N}^{\rm env } _{p\varphi}(\chi'_{12},t)$. }
\end{figure}

\section{Decoherence of the maximally entangled mixed states}

We will analyze the decoherence process of the initially maximally
entangled mixed states of two qubits \cite{Ish00,Mun01,Ver01} on
the example of the Werner states \cite{Wer89} defined to be ($1/3 <
p \le 1$):
\begin{eqnarray}
\hat{\rho}_{p\psi\pm }(0) &=&p|\psi _{\pm}\rangle \langle \psi
_{\pm}|+\frac{1-p}{4} \hat{I}_{1}\otimes \hat{I}_{2},  \label{N36} \\
\hat{\rho}_{p\phi\pm }(0) &=&p|\phi _{\pm}\rangle \langle \phi
_{\pm}|+\frac{1-p}{4} \hat{I}_{1}\otimes \hat{I}_{2}  \label{N37}
\end{eqnarray}
where $|\psi _{\pm}\rangle$ and $|\phi _{\pm}\rangle$ are given by
(\ref{N19}) and (\ref{N23}), respectively, and $\hat{I}_{1,2}$ are
the identity 2x2 matrices. Thus, the Werner states are mixtures of
a MES (Bell state) and the maximally mixed state, given by
$\hat{I}_{1}\otimes \hat{I}_{2}$, which can be interpreted as an
equal incoherent mixture of the four Bell states. It is worth
mentioning that the standard two-qubit Werner state is defined as
$\hat{\rho}_{p\psi-}(0)$ only \cite{Wer89}. This state, given in
terms of the singlet state $|\psi_{-}\rangle$, is invariant if both
qubits are subjected to same unitary transformation, $U\otimes U$.
Nevertheless, by ignoring the $U\otimes U$ invariance but keeping
the same entanglement properties, the standard Werner state is
often generalized (see, e.g., \cite{Wei03,Mun01,Gho01}) to include
mixtures of any MESs, as given by (\ref{N36}) and (\ref{N37}).
Following this convention, we will apply the generalized
definitions of Werner states in our study.

It is easy to show that the concurrences and negativities of the
Werner states are the same and given by
\begin{eqnarray}
C_{p\psi }(0)=C_{p\phi }(0)= \mathcal{N}_{p\psi
}(0)=\mathcal{N}_{p\phi }(0)=(3p-1)/2. \label{N38}
\end{eqnarray}
The Werner states can be considered the MEMSs since their degree of
entanglement cannot be increased by any unitary operations
\cite{Ish00} and they have the maximum of entanglement for a given
linear entropy (and vice versa) \cite{Mun01}. In a special case of
$p=1$, the Werner states go over into the MESs. The evolution of
$\hat{\rho}_{p\psi }(t)$ for the initial Werner state (\ref{N36})
in the lossy nonlinear cavity is described by
\begin{eqnarray}
\hat{\rho}_{p\psi\pm }(t)
&=&\frac{1}{4}\{[(2-g)^{2}-g^{2}p]|00\rangle \langle
00|+g^{2}(1-p)|11\rangle \langle 11|  \nonumber \\
&&\pm 2gp[e^{i(\chi _{1}-\chi _{2})t}|01\rangle \langle
10|+e^{-i(\chi
_{1}-\chi _{2})t}|10\rangle \langle 01|]  \nonumber \\
&&+g[2-g(1-p)](|01\rangle \langle 01|+|10\rangle \langle 10|)\},
\label{N39}
\end{eqnarray}
being independent of the cross-coupling $\chi _{12}$, which
implies that a monotonical decrease of the entanglement occurs
according to
\begin{eqnarray}
C_{p\psi }(t) &=&\max \{0,g
p-g\sqrt{(1-g)(1-p)+\frac{g^{2}(1-p)^{2}}{4}}\},
\nonumber \\
\mathcal{N}_{p\psi }(t) &=&\max
\{0,\sqrt{(1-g)^{2}+g^{2}p^{2}}-\frac{g^2(1-p)}{2} -(1-g)\}.
\label{N40}
\end{eqnarray}
In a special case of $p=1$, the above formulas simplify to
(\ref{N21}) and (\ref{N22}), respectively.  On the other hand, the
evolution of $\hat{\rho}_{p\phi }(t)$ from the initial Werner
state (\ref{N37}) reads as
\begin{eqnarray}
\hat{\rho}_{p\phi\pm }(t) &=&\frac{1}{2}\{(2-2g+x_{p})|00\rangle
\langle 00|
+(g-x_{p})(|01\rangle \langle 01|+|10\rangle \langle 10|)  \nonumber \\
&&\pm p(f|00\rangle \langle 11|+f^{\ast }|11\rangle \langle
00|)+x_{p}|11\rangle \langle 11| \} \label{N41}
\end{eqnarray}
where $x_{p}=(1+p)g^{2}/2$ and $f=g\exp [i(\chi _{1}+2\chi
_{12}+\chi _{2})t]$. Hence, the time evolution explicitly depends
on the cross-coupling $\chi _{12}$, but in such a way that the
concurrence and negativity exhibit the same monotonous decrease
independent of $\chi _{12}$ as follows:
\begin{equation}
C_{p\phi }(t)=\mathcal{N}_{p\phi }(t)=\max
\{0,\frac{g}{2}[g(1+p)-2(1-p)]\} \label{N42}
\end{equation}
In a special case of $p=1$, equation (\ref{N42}) reduces to
$g^{2}$ in agreement with (\ref{N25}). Note that the subscript
$\pm$ in $C_{p\psi }$, $N_{p\psi }$, $C_{p\phi }$, and $N_{p\phi
}$ has been omitted as the functions are independent of the sign
in (\ref{N36}) and (\ref{N37}).

As the last example, let us assume qubits to be initially in the
Werner-like state defined by ($1/3 < p \le 1$):
\begin{equation}
\hat{\rho}_{p\varphi }(0)=p|\varphi \rangle \langle \varphi
|+\frac{1-p}{4} \hat{I}_{1}\otimes \hat{I}_{2}  \label{N43}
\end{equation}
in terms the MES given by (\ref{N26}). The concurrence and
negativity for (\ref{N43}) are equal to $C_{p\varphi
}(0)=\mathcal{N}_{p\varphi }(0)=(3p-1)/2$ being the same as for the
other Werner states. However, its evolution essentially differs
from $\hat{\rho}_{p\psi\pm }(t)$ and $\hat{\rho}_{p\phi \pm}(t) $
by exhibiting oscillations of the entanglement. In detail, it is
described by the density matrix elements
\begin{equation}
\lbrack \hat{\rho}_{p\varphi }(t)]_{ij}=p^{(1-\delta
_{ij})}[\hat{\rho} _{\varphi }(t)]_{ij}  \label{N44}
\end{equation}
given in terms of (\ref{N27}), but with the off-diagonal terms
multiplied by $p$ as $\delta _{ij}$ stands for Kronecker delta. In
a special case of the lossless nonlinear cavity, the entanglement
of the state $\hat{\rho}_{p\varphi }(\gamma =0,t)$ evolves
periodically as follows:
\begin{eqnarray}
C_{p\varphi }(\gamma =0,t) =\mathcal{N}_{p\varphi }(\gamma =0,t)
=\frac{1}{2}\max \{0,p(2|\cos (\chi _{12}t)|+1)-1\} \label{N45}
\end{eqnarray}
which is opposite to the time-independent evolution of the other
Werner states, viz. $\hat{\rho}_{p\psi\pm }(\gamma =0,t)=$
$\hat{\rho}_{p\phi\pm }(\gamma =0,t)=$ const. One can conclude
from (\ref{N45}) (see also figures 3 and 4) that by decreasing
parameter $p$, the entanglement and the time intervals in which
the states are entangled decrease. For the dissipative nonlinear
cavity, the entanglement corresponding to the evolution of
$\hat{\rho}_{p\varphi }(t) $ exhibits decaying oscillations shown
by curves (c) in figures 3 and 4. As in the former section, we are
mainly interested in the envelopes of these oscillations. In the
special case of $p=1$, when the initial Werner-like state goes
over into the Bell-like state, the concurrence and negativity
envelopes are given by (\ref{N31}) and (\ref{N32}), respectively.
By assuming $\chi _{12}\gg \gamma $, an approximate formula for
the $p$-dependent envelopes of the concurrence can be given by
\begin{eqnarray}
\hspace{-1cm} C_{p\varphi }^{\mathrm{env}}(t) \approx \frac{g}{4}
\max \Big\{0,\frac{1}{ \sqrt{3}}
\Big(\sqrt{x_{p}+4p\sqrt{y_{p}}}-2\sqrt{x_{p}-2p\sqrt{y_{p}}}\Big)
+g+p-2\Big\}  \label{N46}
\end{eqnarray}
in terms of $x_{p}=3G^{2}+2Gp+11p^{2}$ and $ y_{p}=3G^{3}
+G^{2}(10+9p)+G(3+14p)+p(9+16p)$ where $G=2-g$. Note that
(\ref{N46}) for $p=1$ is another approximate formula of the
concurrence envelope for the initial MES $|\varphi \rangle $, but
leading to a slightly worse approximation than that given by
(\ref{N31}). For brevity, the lengthy formula for the $p$
-dependent negativity envelope, $\mathcal{N}_{p\varphi
}^{\mathrm{env}}(t)$, generalizing equation (\ref{N32}), is not
presented explicitly here although was used for plotting the
envelope curves (c) in figure 4. By analyzing figures 3 and 4, we
conclude that
\begin{eqnarray}
{C}_{p\varphi }^{{\rm env}}(t_{n})=C_{p\varphi }(\chi
_{12}>0,t_{n}) \geq
{C}_{p\varphi }(\chi _{12}=0,t_{n}),  \nonumber \\
\mathcal{N}_{p\varphi }^{{\rm env}}(t_{n})=\mathcal{N}_{p\varphi
}(\chi _{12}>0,t_{n}) \geq \mathcal{N}_{p\varphi }(\chi
_{12}=0,t_{n}) \label{N47}
\end{eqnarray}
at moments of time $t_{n}\approx n\pi /\chi _{12}$ ($n=1,2,\cdots
$), which means that the decay of entanglement of the initially
Werner-like state (\ref{N43}) in a lossy cavity, can be
periodically retarded by inserting the Kerr nonlinearity in the
cavity.

\section{Discussion and conclusions}

\begin{figure}
\epsfxsize=6cm\centerline{\epsfbox{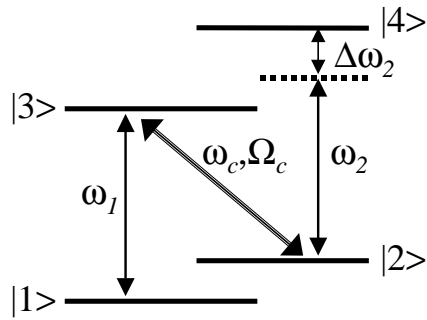}} \caption{Four-level
atomic system for the resonantly enhanced Kerr nonlinearity in the
Schmidt-Imamo\v{g}lu scheme.}
\end{figure}

Let us finally address the question whether the interactions
studied in this paper can be experimentally observable. As
mentioned in Introduction, the conditions assumed in the paper of
the strong Kerr interaction at low light intensities can be
satisfied, e.g., for the EIT schemes as studied theoretically
\cite{Sch96,Ima97,Reb99,Dua00,Hon02,Kua03} and confirmed
experimentally \cite{Hau99,Kan03}.  Schmidt and Imamo\v{g}lu
\cite{Sch96} have proposed a renown EIT scheme where a low density
cloud of cold atoms with the four-level structure, shown in figure
5, exhibits giant resonantly-enhanced nonlinear cross-coupling with
vanishing linear susceptibilities at low intensities. In the
scheme, atoms are placed in a cavity (or double cavity) tuned to
two frequencies: $\omega_1$ of the mode $a_1$ resonant with the
transition $|1\rangle \leftrightarrow|3\rangle$, and $\omega_2$ of
the mode $a_2$ detuned by $\Delta\omega_2$ of the transition
$|2\rangle\leftrightarrow|4\rangle$. The EIT effect for the modes
$a_{1}$ and $a_{2}$ is induced by a classical coupling field of
frequency $\omega_{c}$ resonant with the transition $|2\rangle
\leftrightarrow |3\rangle$. By assuming that $|\omega _{1}-\omega
_{2}| \gg \Delta \omega_{2}$, no nonlinear self-coupling will occur
in the system. Nevertheless, as shown in the preceding sections,
only the cross-coupling changes the entanglement evolution. By
adiabatically eliminating all the atomic levels, Schmidt and
Imamo\v{g}lu have found the real part of the resonantly-enhanced
third-order nonlinear susceptibility, ${\rm Re}(\chi^{(3)})$, to be
given by $|\mu_{13}|^2|\mu_{24}|^2 n_{\rm at}(2\epsilon_0 \hbar^3
\Omega_c^2 \Delta\omega_2 V_{\rm cav})^{-1}$, where $\mu_{ij}$ is
the electric dipole matrix element between the states $|i\rangle$
and $|j\rangle$, $n_{\rm at}$ is the total number of atoms
contained in the cavity of volume $V_{\rm cav}$, $\Omega_c$ is the
coupling-field Rabi frequency, and $\epsilon_0$ is the permittivity
of free space. With the help of expression for ${\rm
Re}(\chi^{(3)})$, it is easy to show that the Kerr nonlinear
cross-coupling is given by \cite{Ima97,Dua00}
\begin{equation}
2\chi_{12} \sim \frac{3|g_{13}|^2|g_{24}|^2}{ \Omega_c^2
\Delta\omega_2}n_{\rm at}\label{N48}
\end{equation}
where $g_{ij}=\mu_{ij}\sqrt{\omega_i/(2\epsilon \hbar V_{\rm
cav})}$ is the coupling coefficient between the atoms and the
cavity mode $a_i$ of frequency $\omega_i$. It is worth stressing
that the above formulas for ${\rm Re}(\chi^{(3)})$ and $\chi_{12}$
are valid under condition that $|g_{13}|^2 n_{\rm
at}/\Omega_c^{2}<1$ required by the applied adiabatic elimination
procedure \cite{Gra98}. The EIT enhanced Kerr-coupling constants
for the Schmidt-Imamo\v{g}lu scheme can be estimated moderately as
$\sim$ 0.2~rad~MHz \cite{Dua00} or, by putting the stringent limit
on the required cavity parameters \cite{Gra98}, as $\sim$
100~rad~MHz \cite{Ima97}. In our numerical analysis, we have chosen
$\chi_{12}= 20$~rad~MHz. A typical cavity decay rate obtainable in
current experiments is of the order $\sim$ 4~rad~MHz, which is 5
times smaller than the value of $\chi_{12}$ chosen for plotting
figures 1--4. This estimation is less stringent than that given in
Ref. \cite{Ima97}. It is worth noting that the EIT leads to
remarkable light-speed reduction \cite{Hau99}, which enables
reduction of the cavity decay rate in the Schmidt-Imamo\v{g}lu
setup with the same finesse mirrors. We have analyzed decays within
times $<80$ns, which are fairly shorter than dephasing time ($\sim
9\mu$s) for atom cloud measured in the Hau et al. experiment
\cite{Hau99} but longer, for the obvious reasons, than the
evolution times ($\sim 8$ns) in the quantum non-demolition scheme
of Duan et al. \cite{Dua00}.

In conclusion, we have analyzed evolution of two optical modes in
qubit states interacting via a Kerr nonlinearity  in a lossy cavity
modelled by dissipative coupled nonlinear oscillators being
initially in the maximally entangled pure or mixed states. We have
found that for the initial Bell ($|\psi_{\pm}\rangle$ and
$|\phi_{\pm}\rangle$) or Bell-like ($|\varphi\rangle$) states, the
decay of the concurrence, or equivalently the entanglement of
formation, is the slowest for $|\psi_{\pm}\rangle$ and the fastest
for $|\phi_{\pm}\rangle$, while the decay of the negativity, or
equivalently the PPT-entanglement cost, is the slowest for
$|\varphi\rangle$ (if the nonlinearity parameter is much greater
than the damping constants) and the fastest for
$|\psi_{\pm}\rangle$. Thus, we have provided simple analytical
examples of states differently ordered by the concurrence and
negativity. These seemingly inconsistent results are physically
meaningful as discussed in, e.g., Refs. \cite{Eis99,Zyc99,Wei03}
and proved in general terms by Virmani and Plenio \cite{Vir00}.
Nevertheless, to our knowledge, our analysis is the first
demonstration of the relativity of the entanglement measures as a
result of a physical process. Moreover, we have also studied
decoherence of the initial maximally entangled mixed states on the
example of three kinds of Werner(-like) states as related to the
different Bell(-like) states $|\psi_{\pm}\rangle$,
$|\phi_{\pm}\rangle$ and $|\varphi\rangle$. Our analytical and
numerical results show the differences and similarities of the
negativity and concurrence decays of the Werner{(-like)} states in
comparison to the Bell(-like) states.

We have demonstrated that by inserting medium with the Kerr
nonlinearity, described by Hamiltonian (\ref{N10}), into the lossy
cavity, evolution of the initial Bell states $|\psi_{\pm}\rangle$
or $|\phi_{\pm}\rangle$ and the corresponding Werner states is
changed but in such a way that the entanglement decays in the same
manner as without the nonlinear medium. However, if the qubits
placed in a lossy cavity are initially in the Bell-like state
$|\varphi\rangle$ or the corresponding Werner-like state, the loss
of entanglement can be periodically delayed (partially recovered)
by inserting medium with the Kerr nonlinearity.

\section*{Acknowledgments}

The author thanks Jens Eisert, Nobuyuki Imoto, Masato Koashi,
Wies\l{}aw Leo\'nski, Yu-xi Liu, \c{S}ahin K. \"{O}zdemir, and
Ryszard Tana\'{s} for stimulating discussions.

\end{document}